\documentclass[doublecol]{epl2} 

\usepackage{amsmath}
\usepackage{url,hyperref,lineno,microtype,subcaption}

\title{The hidden  fluctuation-dissipation theorem for growth}
\shorttitle{Title} 

\author{M\'{a}rcio S. Gomes-Filho\inst{1} \and Fernando A. Oliveira\inst{2}}
\shortauthor{Gomes-Filho \etal}

\institute{                    
  \inst{1} Instituto de F\'{i}sica, Universidade de Bras\'{i}lia, Bras\'{i}lia-DF, Brazil\\
  \inst{2} Instituto de F\'{i}sica, Universidade Federal da Bahia, Campus Universit\'{a}rio da Federa\c{c}\~{a}o,
 Rua Bar\~{a}o de Jeremoabo s/n, 40170-115, Salvador-BA, Brazil
}

\abstract{In a stochastic process, where noise is always present,  the fluctuation-dissipation theorem (FDT)  becomes one of the most important tools in statistical mechanics and, consequently, it appears everywhere. Its major utility is to provide a simple response to study certain processes in solids and fluids. However, in many situations we are not talking about a FDT, but about the noise intensity.  For example, noise has enormous importance in diffusion and growth phenomena. Although we have an explicit FDT for diffusion phenomena, we do not have one for growth processes where we have a noise intensity. We show  that  there is a hidden FDT for the growth phenomenon, similar to the diffusive one. Moreover, we show that growth with correlated noise presents as well a similar form of FDT. We also call attention to the hierarchy within the theorems of  statistical mechanics and how this explains the violation of  the FDT in some phenomena.
}

\begin{document}

\maketitle

\textit{ Introduction.\textemdash} The  era of explicit stochastic process in physics started with Langevin's analysis of the Brownian motion~\cite{Brown28,Brown28a,Einstein1905, Einstein56, Langevin08,Vainstein06,Nowak17,Oliveira19}, by considering the equation of motion for a particle moving in a fluid as~\cite{Langevin08}: 
\begin{equation}
m\frac{dv(t)}{dt}=-m\gamma v(t)+f(t),
\label{L}
\end{equation}
where $m$ is the mass of the particle and $\gamma$ is the friction. The ingenious and elegant proposal was to modulate the complex interactions between  particles, considering all interactions as two main forces.
The first contribution represents a frictional force, $-m\gamma v$, where the characteristic time scale is $\tau=\gamma^{-1}$ while the second contribution comes from a stochastic force, $f(t)$, with time scale $\Delta t \ll \tau $, which is related with the random collisions between the particle and the fluid molecules. 

The fluctuating force $f(t)$, in eq.~(\ref{L}), obeys the following conditions:\\
$(i)$ the mean force due to the random collisions on the particle is zero,
\begin{equation}
\label{fmed}
\langle f(t)\rangle =0,
\end{equation}
$(ii)$ there is no correlation between the initial particle velocity and the random force,
\begin{equation}
\label{fv}
\langle f(t)v(0)\rangle =0 
\end{equation}
\noindent and $(iii)$ the fluctuating forces at different times $t$ and $t'$ are uncorrelated
\begin{equation} 
\label{whitenoise1}
\langle f(t)f(t')\rangle= 2\rho \delta(t-t'),
\end{equation}
where $\rho$ is the noise intensity.  A solution for (\ref{L}) is \cite{Oliveira19}
\begin{equation}
 v(t)=v(0)+\frac{1}{m}\int_0^t f(s) \exp[-\gamma(t-s)] ds,
\end{equation}
which gives $<v(t)>=<v(0)> \exp(-\gamma t)$ and 
\begin{equation}
\label{sigv}
  \langle v^2(t) \rangle = \langle v^2(0) \rangle \exp(-2\gamma t)+\frac{\rho}{m^2\gamma}\left[1-\exp(-2\gamma t)\right].
\end{equation}
In order to get  the above relation we have used the conditions $(i)$, $(ii)$ and $(iii)$. 
Now, considering that $ \langle v^2(t \rightarrow \infty) \rangle = \langle v^2 \rangle_{eq}$ as $t \rightarrow \infty$, we get for the FDT:
\begin{equation} 
\label{FDT}
\langle f(t)f(t')\rangle= 2m^2\gamma \langle v^2 \rangle_{eq} \delta(t-t').
\end{equation}
Equation (\ref{FDT}) is the FDT in its simplest form, i.e. a relation between the fluctuating force $f(t)$ and the dissipation $\gamma$.
Note that the equipartition theorem  states that $\langle v^2 \rangle_{eq}= k_BT/m$, where  $k_B$ is the Boltzmann constant and $T$ the absolute temperature. 
It is noteworthy that the artificial separation between the stochastic force and the dissipative force in the eq.~(\ref{L}) now disappears. An important relation was pointed out between them. This main relation is the FDT. We show below that there is a similar relation in growth phenomena. 
 
 \textit{The growth fluctuation-dissipation theorem.\textemdash}  Since growth phenomena are widely found in many dynamical processes in biology, chemistry and physics, the surface dynamics has become an intense research topic in the last decades~\cite{Hansen00,Barabasi95,Edwards82,Kardar86}. Although these systems are complex in nature, they can be modeled in simple ways. For instance, the statistical character of the evolutionary process leads us to an inhomogeneous surface, which can be  described by a height $h(\vec{x},t)$, being  $\vec{x}$ the position in $d$ dimensional space and $t$ the time. Two quantities play an important role in growth, the average height, $\langle h \rangle$, and  the  standard deviation 
\begin{equation}
w(t)= \left[ \langle h^2(t) \rangle - \langle h(t) \rangle^2\right]^{1/2},
\label{wt}
\end{equation}
which is named as roughness or the surface width. The average here is obtained over the sites. It should be mentioned that, again, the fluctuation is the main physical quantity, and we would say that, in statistical physics, it is second only to the diffusion dispersion, eq.~(\ref{sigv}), since many important phenomena and processes have been associated with it.

The general characteristics of the growth dynamics were observed through some analytical, experimental and computational results~\cite{Barabasi95}. For many growth processes, the roughness, $w(t)$, increases with time until reaches a saturated roughness $w_s$, i.e., $w(t \rightarrow \infty)=w_s$. We can summarize the time evolution of all regions as following:
 \begin{equation}
\label{Sc1}
w(t,L)=
\begin{cases}
 ct^\beta , &\text{ if~~ } t <<t_\times\\
 w_s \propto L^\alpha, &\text{ if~~ } t >> t_\times,\\
\end{cases}
\end{equation}
with $t_{\times} \propto L^z$. The dynamical exponents satisfy the general scaling relation:
\begin{equation}
\label{z}
z=\frac{\alpha}{\beta}.
\end{equation}
Different methods have been proposed to understand this rich phenomenon. Here, we will focus only on processes that saturate. For example, the attempt to describe the height evolution leads us to some dominant type of Langevin's equations such as Edwards-Wilkinson equation (EW)~\cite{Edwards82}:
\begin{equation}
\label{EW}
\dfrac{\partial h(\vec{x},t)}{\partial t}=\nu \nabla^2 h(\vec{x},t)  + \xi(\vec{x},t),
\end{equation}
and the  Kardar-Parisi-Zhang (KPZ) equation~\cite{Kardar86}:
\begin{equation}
\label{KPZ}
\dfrac{\partial h(\vec{x},t)}{\partial t}=\nu \nabla^2 h(\vec{x},t) +\dfrac{\lambda}{2}[\vec{\nabla}h(\vec{x},t)]^2 + \xi(\vec{x},t),
\end{equation}
where the parameters $\nu$ (surface tension) and $\lambda$ are related to the Laplacian smoothing and the tilt mechanism, respectively. The stochastic process is characterized by the noise, $\xi(\vec{x},t)$, which is defined as a simple form, a white noise:
 
\begin{equation}
\label{xi}
\left\langle \xi(\vec{x},t) \xi(\vec{x'},t')\right\rangle = 2D\delta^d(\vec{x}-\vec{x'})\delta(t-t'),
\end{equation}
where $D$ is here the noise intensity. For the KPZ universality class,  we have the Galilean invariance~\cite{Kardar86}:
\begin{equation}
\label{GI}
\alpha+z=2.
\end{equation}

In this way, the KPZ equation, eq.~(\ref{KPZ}), is a general nonlinear stochastic differential equation, which can characterize the growth dynamics of  many different systems~\cite{Mello01,Odor10,Merikoski03,Takeuchi13,Almeida17,Reis05,Almeida14,Rodrigues15,Alves16,Carrasco18}. Despite all effort, finding an analytical solution of the KPZ equation~(\ref{KPZ}) is not an easy task~\cite{Dasgupta96,Dasgupta97,Torres18,Wio10a,Wio17,Rodriguez19} and we are still far from a satisfactory theory for the KPZ equation, which makes it one of the most difficult problems in modern mathematical physics~\cite{Bertine97,Baik99,Prahofer00,Dotsenko10,Calabrese10,Amir11,Sasamoto10,Doussal16,Hairer13}, and probably one of the most important problem in nonequilibrium statistical physics.
 
 It should be noted that there is a gap in our understanding of what the FDT  for the growth process is, since the intensity $D$ is not directly connected with other properties of the force as in the Langevin equation (LE). To overcome this, let us remember that in the LE the equilibrium is reached when we apply the equipartition theorem,  eq.~(\ref{sigv}), which establishes that we have 
$\langle v^2(t \rightarrow \infty) \rangle \equiv v_s^2=k_BT/m$, where $v_s^2$ is the final squared average velocity or the ``saturated'' velocity, that leads to eq.~(\ref{FDT}). 

For the EW and KPZ equation we have as well that the noise  increases $h$ and $w(t)$ linearly with time, and the term $\nu \nabla^2 h(\vec{x},t)$, which is always opposite to roughness decreases it. For  $1+1$ dimensions we have the exact results
 \begin{equation}
\label{Ws}
w_s=\sqrt{\frac{cD}{\nu}L},
\end{equation}
being  $c=\frac{1}{4 \pi^2 }$ for Edwards-Wilkinson~\cite{Edwards82} and $c=\frac{1}{24}$ for the single step (SS) model~\cite{Krug92,Krug97}. For growth, we do not have an equipartition theorem, but we have  $w^2(t \rightarrow \infty) = w_s^2$, thus we replace $D$ by the above value to obtain
\begin{equation}
\label{GFDT}
\left\langle \xi({x},t) \xi({x'},t')\right\rangle = 2b\nu w_s^2\delta({x}-{x'})\delta(t-t'),
\end{equation}
with $b=1/(cL)$. Therefore, the parameter $D$ in eq.~(\ref{xi}) is not only the noise intensity but it is also related to $\nu$, in the above equation, which leads to a fluctuation-dissipation theorem (FDT). Note the similarity between eq.~(\ref{GFDT}) and eq.~(\ref{FDT}). Also, since the noise and the surface tension in the Edwards-Wilkinson equation have their origin in the flux, the  separation between them  is artificial, consequently,  the above discussion restore the lost link.  

It is important to note that the only difference between the Edwards-Wilkinson and the KPZ FDT is in the constant $c$.  Considering that the growth phenomenon is much more complex than diffusion, this is an impressive result.  Moreover, $c$ depends only on the universality class, particularly for KPZ, it corresponds to a large number of different models.   Note as well that the dispersion of the velocity is proportional to $T$ while the roughness is proportinal to $L$, which means that $<v^2>_{eq} \propto T$ and $w_s^2 \propto L$, respectively. Therefore, the size $L$ is the ``ingredient'' that allows disorder in growth phenomena.

Now, if we add a force in the Langevin equation, linear or not, it does not affect the FDT. The same for the EW equation: if we add, for example, $\dfrac{\lambda}{2}[\vec{\nabla}h(\vec{x},t)]^2,$ transforming it in the KPZ equation, it does not alter the FDT. It is noteworthy that for the SS model the height difference between next neighbors is $\pm1$, then the $[\vec{\nabla}h(\vec{x},t)]^2$ is just a constant,  which for a stronger reason it will not affect the FDT, except for the constant $b$ above.  Consequently, there is an implicit FDT for the growth process, similar to the Langevin equation. This was the missing part that needed to be clarified. Besides, note that the  equation~(\ref{GFDT}) depends on Periodic Boundary Conditions (PBC) because both Edwards-Wilkinson~\cite{Edwards82} and Krug {\it et al.}~\cite{Krug92,Krug97} result into eq.~(\ref{Ws})  was obtained using  PBC.

\begin{figure}[tbp]
	\begin{center}
 	 \includegraphics[width=1\columnwidth]{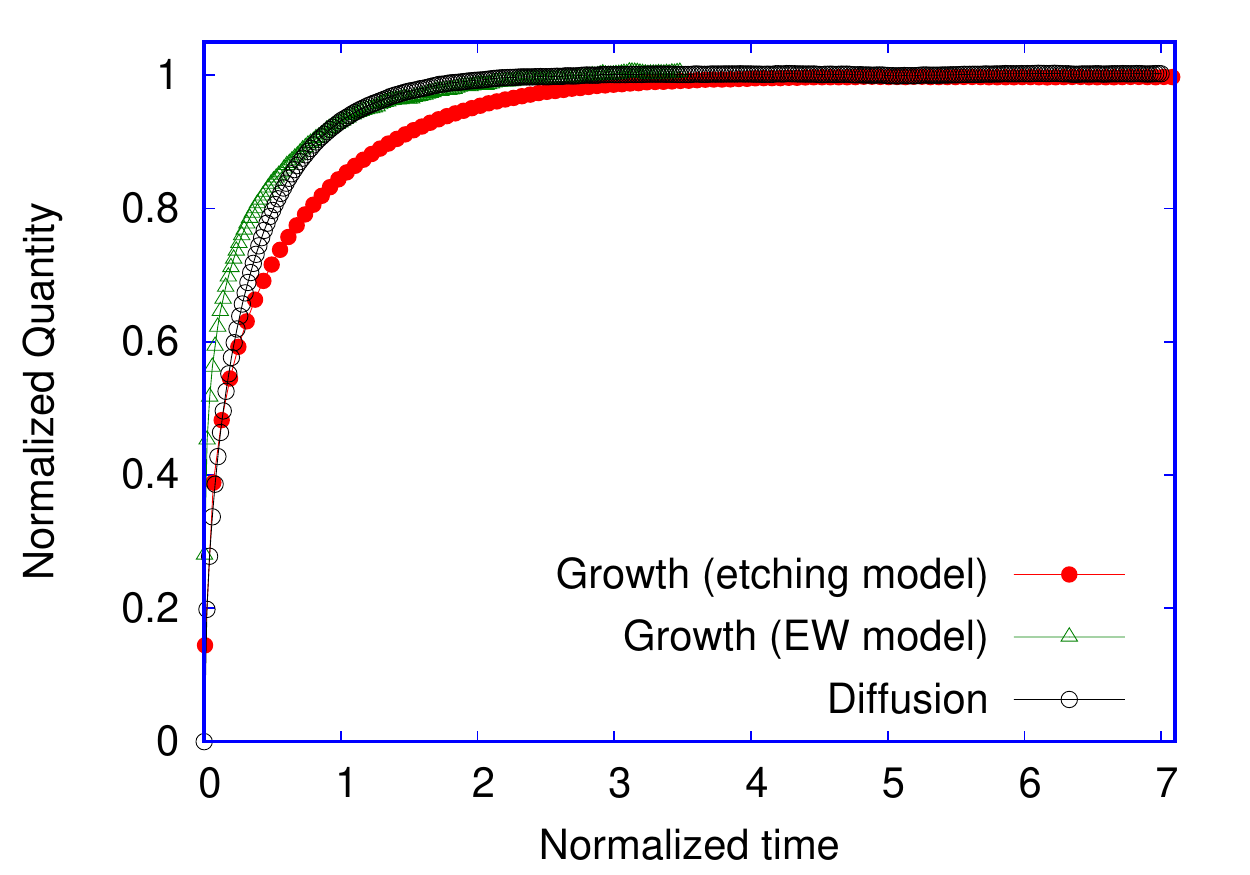}	
	\caption{\label{fig1} Dispersions as a function of time. For diffusion, we have the mean squared velocity in units of  $\sqrt{k_BT/m}$; For growth, the normalized roughness, $w(t)/w_s$, as function of their normalized time for the KPZ universality class (etching model) and for the Edwards-Wilkinson universality class (EW model).}	
	\end{center}
\end{figure}

In this way, we show in Fig~(\ref{fig1}) the evolution of the dispersions  with time. In the upper curve we exhibit the  mean square velocity $\tilde{v}=\sqrt{\langle v^2(t ) \rangle }$ in units of the ``saturated'' or mean squared  equilibrium velocity  $\sqrt{k_BT/m}$, and time in units of normalized time $t/\tau$ with $\tau=\gamma^{-1}$. The curve was obtained by solving eq.~(\ref{L}), with the initial condition $v(t=0)=0$, and averaging over $1\times 10^6$ numerical experiments.

Simulations of the growth equations are usually more hard~\cite{Takeuchi13}, thus we use cellular automaton model in the same class of universality of the growth stochastic equation. In this way, the middle curve is for the Edwards-Wilkinson model,  we plot the  roughness, $w(t)$, normalized to the proper units, i.e  $w(t)/w_s$, as a function of normalized time $t/t_\times$. In order to model the EW equation we shall use the SS model. The SS model has gained a lot of importance over the years~\cite{Krug92,Krug97,Derrida98,Meakin86,Daryaei20}: firstly,  because it was proved to be a KPZ model and secondly   due to its connection with other models, such as the asymmetric simple exclusion process~\cite{Derrida98}, the six-vertex model~\cite{Meakin86,Gwa92,Vega85}, and the kinetic Ising model~\cite{Meakin86, Plischke87}.  Finally, our interest here is because this model, in particular, can become an EW model.  The SS model is defined in such way that the height difference between two neighbors heights $\eta=h_i-h_j$ is just $\eta=\pm 1$. Consequently, it is easily associated  to the Ising model. For example, for $1+1$ dimensions the initial conditions for the height of the site $i$ are of the form 
$h_i(0)=(1+(-1)^i)/2$. Now, let us consider in $1+1$ dimensions a chain of size $L$ and we define the space $\Omega=1,2,3,\cdots,L-1,L$. First, we select a site $i$ randomly, then we compare its height  with its neighbors, and we apply the following rules:
\begin{enumerate}
\item At time $t$, randomly choose a site $i\in{\Omega}$;
\item If $h_i(t)$ is a minimum, then $h_i(t+\Delta t)=h_i(t)+2$, with probability $p$;
\item If $h_i(t)$ is a maximum, then $h_i(t+\Delta t)=h_i(t)-2$, with probability $q$.
\end{enumerate} 
With the above rules we can generate the dynamics of the SS model for $d+1$ dimensions. For $1+1$ dimensions its properties have been well studied~\cite{Krug92,Krug97,Derrida98}, and we can obtain analytically results such as
\begin{equation}
\label{lambda0}
\lambda =p-q.
\end{equation}
Note that for $p=q$, it becomes the EW model.  Changing the probabilities we can get a lot of relevant information~\cite{Daryaei20}. Now, we start to simulate the model with $p=q=1/2$, i.e. $\lambda=0$ (EW model),  with $L=128$ and the initial condition $h_i(0)=(1+(-1)^i)/2$.  We use the above rules, periodic boundary conditions and we average over $1 \times 10^4$ numerical experiments to obtain the middle curve with $w_s = 3.161(1)$ and $t_x = 514.5(9)$.

The lower curve is the  roughness $w(t)$, normalized to proper units, i.e  $w(t)/w_s$, as a function of normalized time $t/t_\times$, which was obtained from the $1+1$ etching model.  The etching model~\cite{Mello01,Gomes19,Rodrigues15,Alves16} is a stochastic cellular automaton that simulates the surface erosion due to action of an acid. This model is proposed as simple as possible in order to capture the essential physics, which considers that the probability of removing a cell is proportional to the number of the exposed faces of the cell (an approximation of the etching process). First, we randomly select a site $i$ with a certain height, $h_i$, and then we compare it with one of its nearest neighbor $j$.  This is similar to the SS model, but the rules change: if $h_{j} > h_i$, it is reduced to the same height as $h_i$, which means that the height of the surface decreases at each step. The main algorithm steps are summarized as follows:
\begin{enumerate}
\item At time $t$, randomly choose a site $i\in{V}$;
\item If $h_{j}(t) >  h_i(t)$, then  $h_{j}(t+\Delta t)=h_i(t)$;
\item Consider  $h_i(t+\Delta t)=h_i(t)-1.$
\end{enumerate}
This stochastic cellular automaton has  the advantage of allowing  us to understanding a corrosion process, and at same time to study the KPZ growth process.  Note that the rule $1)$ is commom to all growth models, it is equivalent to the random term $ \xi(\vec{x},t)$ in the KPZ equation~(\ref{KPZ}). The interaction between the neighbors is equivalent to diffusive ones, and  the nonlinear term represents the lateral growth. 
The etching model belongs to the KPZ universality class~\cite{Alves16,Gomes19}. Now we use these rules to simulate this model. We take the initial condition as $h_i(t=0)=0$, and we use periodic boundary conditions.   We take $L=128$ and we average over $4 \times 10^6$ numerical experiments, and we obtain  $w_s = 7.56(2)$ and $t_x = 163.95(3)$.   As we can see, all curves grow with time and afterwards all of them are equilibrated or thermalized.

Consequently, the saturation phenomenon is a competition between noise and smoothing mechanism, and since they are part of the same flux in the growth equations, the separation between them is artificial as it is in the Langevin equation.  The above discussion help us to restore the lost link. Finally, note that the final result for the FDT, for both eq.~(\ref{FDT}) and eq.~(\ref{GFDT}), does not depend on the initial conditions. However, the eq.~(\ref{GFDT}) depends on periodic boundary conditions.


\textit{Correlated noise.\textemdash} As exposed above, the FDT has an important place in statistical mechanics. Not only because it is a basic statement, but also because it allows us to obtain important measured quantities, such as susceptibility, the light scattering cross section, the neutron scattering intensity, diffusion, surface roughness in growth, and so on. We make now  a generalization of the previous result for correlated noise.


 A natural extension of the Langevin formalism is the Mori equation~\cite{Mori65,Mori65a,Kubo91,Lee83,Lee82,Lee84,Florencio85} for the operator $\hat{O}(t)$ as 
\begin{equation}
\label{GLE}
\frac{  d\hat{O}(t)}{dt} =-\int_0^t \Gamma(t-s) \hat{O}(s)ds+ f(t).
\end{equation}
 Now, the fluctuating force, $f(t)$, obeys the following conditions:\\
$(i)$ the mean force is zero,
\begin{equation}
\label{fmed2}
\langle f(t)\rangle =0,
\end{equation}
$(ii)$ there is no correlation between the initial value for the operator and the random force,
\begin{equation}
\label{fv2}
\langle f(t)\hat{O}(0)\rangle =0
\end{equation} 
and $(iii)$ the fluctuating forces at different times $t$ and $t'$ are correlated as
\begin{equation} 
\label{colornoise1}
\langle f(t)f(t')\rangle= \Gamma(t-t').
\end{equation}

The basic difference from the Langevin's formalism is that we now have a correlation and the above relation defines a memory function $\Gamma(t)$, and the Mori formalism is a Quantum formalism. Note that for the particular case $\Gamma(t)=2 \gamma \delta(t)$ we return to the Langevin's formalism. Now we use the above conditions for the noise to obtain a self-consistent equation for the correlation function $R(t)$, 
\begin{equation}
R(t)=\frac{\langle \hat{O}(t)\hat{O}(0) \rangle}{\langle \hat{O}(0)^2 \rangle},
\end{equation}
namely
\begin{equation}
\label{self_consistent}
\frac{d R(t)}{d t}=-\int_0^t R(t-t')\Gamma(t')\,d t'.
\end{equation}

The determination of  correlation function is fundamental for the determination of the dynamics. Thus, the equation~(\ref{self_consistent}) is an important part of the theory, which means,  given $\Gamma(t)$ we can get $R(t)$ and then all dynamics. For the Mori equation, the correlation function has been subject of intense investigation~\cite{Lee82,Lee83,Lee84,Florencio85,Morgado02,Costa03,Lapas07,Lapas08,Florencio20}. Correlated phenomena yield  very peculiar forms of anomalous relaxation~\cite{Vainstein06,Vainstein06a,Kohlrausch54,Kohlrausch63,Khinchin49,Lee07a, Weron10, Dybiec12,Ferreira12,Lapas15}.

The equivalent of that for the growth phenomena is to consider a correlated noise in the asymptotic form~\cite{Barabasi95,Lam92}:
\begin{equation}
\label{GFDTC1}
\left\langle \xi(x,t) \xi(x',t')\right\rangle  \propto |x-x'|^{2\phi-1} |t-t'|^{2\theta-1},
\end{equation}
for the $1+1$ KPZ equation. Here $0<\phi<0.5$ and $0<\theta<0.5$ are non-universal exponents. For small values of $\phi$ and $\theta$ the growth exponents are the same as those of the uncorrelated, or strong local correlation~\cite{Lam92,Medina89}. However,  for $\theta \approx 0.45$, we have  $\alpha  \approx 1.0$~\cite{Lam92}. On the other hand, for the particular case

\begin{equation}
\label{GFDTC2}
\left\langle \xi(x,t) \xi(x',t')\right\rangle = 2D\delta({x}-{x'})|t-t'|^{2\theta-1},
\end{equation}
which is more easy to connect with anomalous diffusion. For example, in this situation $\alpha  \neq 1/2$ and $w_s \propto (D/\nu)^\alpha$, and thus  the eq.~(\ref{GFDTC2}) becomes
\begin{equation}
\label{GFDTC3}
\left\langle \xi(x,t) \xi(x',t')\right\rangle = C_1\nu w_s^{1/\alpha} \delta({x}-{x'})|t-t'|^{2\theta-1},
\end{equation}
where $C_1$ and  $\alpha$ are not universal constants, i.e. they depend on $\theta$. Therefore, the  eq.~(\ref{GFDTC3}) is not universal as the  eq.~(\ref{GFDT}).

\textit{Violation of the FDT: the hierarchy.\textemdash} The violation of the FDT is a very common phenomenon. For example it happens in  ballistic diffusion~\cite{Costa03,Lapas07,Lapas08}, and   in KPZ \cite{Kardar86,Rodriguez19} for $d>1$. The violation of the FDT was also observed in structural glass~\cite{Grigera99,Ricci-Tersenghi00,Crisanti03,Barrat98,Bellon02,Bellon05}, in proteins~\cite{Hayashi07}, and  in mesoscopic radioactive heat transfer as well~\cite{Perez-Madrid09,Averin10}. Thus, an explicit  FDT for the $d+1$ KPZ equation, such as the above one for $1+1$ dimensions goes beyond our objective here.  However, due to the importance of the FDT to understand why it fails is a must. The first step was to realize that there is a hierarchy in the major theorems of statistical physics. 
First, note that if
\begin{equation}
 \label{MC}
  \lim_{t \rightarrow \infty}R(t)=0,
 \end{equation}
namely, the mixing condition (MC), holds, the system forgets  its initial conditions~\cite{Khinchin49,Lee07a} and then reaches a full equilibrium. The Khinchin theorem~\cite{Khinchin49,Lee07a} establishes that if (\ref{MC}) holds then ergodicity holds as well.
In anomalous diffusion the violation of the FDT has been associated with the violation of the mixing condition~\cite{Costa03,Lapas07,Lapas08}, which means
\begin{equation}
 \label{NMC}
  \lim_{t \rightarrow \infty}R(t)=\kappa 
 \end{equation}
with $\kappa \neq 0$, which was denominated as the non-ergodic factor \cite{Costa03}. The sequence of works~\cite{Costa03,Lee07a,Lapas07,Lapas08} was very important because it established  in an exact and analytical way the violation of the FDT and how it happens. The hierarchy is simple, the Khinchin's theorem is more important than ergodicity, which is more important than the FDT. Unfortunately, we did  not  achieve such relation for the $d+1$ KPZ equation.

\textit{Last remarks-\textemdash} The fluctuation is a very common phenomenon and, as such, it is universal. The ``dissipation'', however, is more subtle. In  diffusive motion is easily to accept because it is  the friction force that we are used to.
In the growth phenomenon, it appears as a ``friction'' to the roughness, which means that, since the noise creates local irregularities, the smoothing mechanism  acts with a ``frictional force'' proportional to the local curvature, but with an opposite sign to it, which reduces the roughness. The process continues until a saturated roughness $w_s$ is reached.

\textit{Conclusion.\textemdash} In this work,  we start discussing the elegant formalism of Langevin to describe the  Brownian motion, where fluctuation-dissipation theorem (FDT) was explicitly stated for the first time~\cite{Langevin08}. The FDT establishes a clear connection between noise intensity (strength of the random force) and dissipation for a system in thermal equilibrium. 
 Next, we present some types of Langevin equations, stochastic partial differential equations,  such as the Edwards-Wilkinson  and the Kardar-Parisi-Zhang equation in $1+1$ dimensions to describe the growth dynamics.  From this, we were  able to show  that  there is a hidden FDT for the growth phenomenon, similar to the diffusive one, in the sense that the parameter $D$ (noise intensity) that increases the roughness is directly related to $\nu$ (surface tension),which decreases it.  This gives us explicitly a fluctuation-dissipation theorem. Furthermore, we note that different classes of universality exhibit the same behavior,  with distinction only in the non-dimensional constant. We  also extended our discussion to systems with correlated noise.  We expect that this work will stimulate new investigations in both correlated and higher dimension systems.

 

\acknowledgments
This work was supported by the Conselho Nacional de Desenvolvimento Cient\'{i}fico e Tecnol\'{o}gico (CNPq), Grant No.  CNPq-312497/2018-0  and the Funda\c{c}\~ao de Apoio a Pesquisa do Distrito Federal (FAPDF), Grant No. FAPDF- 00193-00000120/2019-79. (F.A.O.).

\end{document}